\documentclass[aps,prd,superscriptaddress,preprint,preprintnumbers,tightenlines,nofootinbib,floatfix]{revtex4-1}

\usepackage{graphicx} 
\usepackage{dcolumn}  
\usepackage{bm}       

\newcommand{\gev}{\ \rm GeV}

\newcommand{\dspa}{D^+_{\mathrm s}}
\newcommand{\dsma}{D^-_{\mathrm s}}
\newcommand{\mreca}{M_{\mathrm{rec}}}
\newcommand{\noa}{N_{\mathrm{obs}}}
\newcommand{\mads}{M_{\mathrm{Ds}}}

\begin{document}

\title{\boldmath Search for the decay $\dspa\to\omega e^+\nu$ }
\preprint{CLNS 11/2077}  
\preprint{CLEO 11-04}    

\author{L.~Martin}
\author{A.~Powell}
\author{G.~Wilkinson}
\affiliation{University of Oxford, Oxford OX1 3RH, UK}
\author{J.~Y.~Ge}
\author{D.~H.~Miller}
\author{I.~P.~J.~Shipsey}
\author{B.~Xin}
\affiliation{Purdue University, West Lafayette, Indiana 47907, USA}
\author{G.~S.~Adams}
\author{B.~Moziak}
\author{J.~Napolitano}
\affiliation{Rensselaer Polytechnic Institute, Troy, New York 12180, USA}
\author{K.~M.~Ecklund}
\affiliation{Rice University, Houston, Texas 77005, USA}
\author{J.~Insler}
\author{H.~Muramatsu}
\author{C.~S.~Park}
\author{L.~J.~Pearson}
\author{E.~H.~Thorndike}
\affiliation{University of Rochester, Rochester, New York 14627, USA}
\author{S.~Ricciardi}
\affiliation{STFC Rutherford Appleton Laboratory, Chilton, Didcot, Oxfordshire, OX11 0QX, UK}
\author{C.~Thomas}
\affiliation{University of Oxford, Oxford OX1 3RH, UK}
\affiliation{STFC Rutherford Appleton Laboratory, Chilton, Didcot, Oxfordshire, OX11 0QX, UK}
\author{M.~Artuso}
\author{S.~Blusk}
\author{R.~Mountain}
\author{T.~Skwarnicki}
\author{S.~Stone}
\author{L.~M.~Zhang}
\affiliation{Syracuse University, Syracuse, New York 13244, USA}
\author{G.~Bonvicini}
\author{D.~Cinabro}
\author{A.~Lincoln}
\author{M.~J.~Smith}
\author{P.~Zhou}
\author{J.~Zhu}
\affiliation{Wayne State University, Detroit, Michigan 48202, USA}
\author{P.~Naik}
\author{J.~Rademacker}
\affiliation{University of Bristol, Bristol BS8 1TL, UK}
\author{D.~M.~Asner}
\altaffiliation[Now at: ]{Pacific Northwest National Laboratory, Richland, WA 99352}
\author{K.~W.~Edwards}
\author{K.~Randrianarivony}
\author{G.~Tatishvili}
\altaffiliation[Now at: ]{Pacific Northwest National Laboratory, Richland, WA 99352}
\affiliation{Carleton University, Ottawa, Ontario, Canada K1S 5B6}
\author{R.~A.~Briere}
\author{H.~Vogel}
\affiliation{Carnegie Mellon University, Pittsburgh, Pennsylvania 15213, USA}
\author{P.~U.~E.~Onyisi}
\author{J.~L.~Rosner}
\affiliation{University of Chicago, Chicago, Illinois 60637, USA}
\author{J.~P.~Alexander}
\author{D.~G.~Cassel}
\author{S.~Das}
\author{R.~Ehrlich}
\author{L.~Gibbons}
\author{S.~W.~Gray}
\author{D.~L.~Hartill}
\author{B.~K.~Heltsley}
\author{D.~L.~Kreinick}
\author{V.~E.~Kuznetsov}
\author{J.~R.~Patterson}
\author{D.~Peterson}
\author{D.~Riley}
\author{A.~Ryd}
\author{A.~J.~Sadoff}
\author{X.~Shi}
\author{W.~M.~Sun}
\affiliation{Cornell University, Ithaca, New York 14853, USA}
\author{J.~Yelton}
\affiliation{University of Florida, Gainesville, Florida 32611, USA}
\author{P.~Rubin}
\affiliation{George Mason University, Fairfax, Virginia 22030, USA}
\author{N.~Lowrey}
\author{S.~Mehrabyan}
\author{M.~Selen}
\author{J.~Wiss}
\affiliation{University of Illinois, Urbana-Champaign, Illinois 61801, USA}
\author{J.~Libby}
\affiliation{Indian Institute of Technology Madras, Chennai, Tamil Nadu 600036, India}
\author{M.~Kornicer}
\author{R.~E.~Mitchell}
\affiliation{Indiana University, Bloomington, Indiana 47405, USA }
\author{D.~Besson}
\affiliation{University of Kansas, Lawrence, Kansas 66045, USA}
\author{T.~K.~Pedlar}
\affiliation{Luther College, Decorah, Iowa 52101, USA}
\author{D.~Cronin-Hennessy}
\author{J.~Hietala}
\affiliation{University of Minnesota, Minneapolis, Minnesota 55455, USA}
\author{S.~Dobbs}
\author{Z.~Metreveli}
\author{K.~K.~Seth}
\author{A.~Tomaradze}
\author{T.~Xiao}
\affiliation{Northwestern University, Evanston, Illinois 60208, USA}

\collaboration{CLEO Collaboration}
\noaffiliation


\date{\today}

\begin{abstract}
We present the first search for the decay $\dspa\to \omega e^{+}\nu$ to test the four-quark content of the $\dspa$ and the 
$\omega$-$\phi$ mixing model for this decay. 
We use $586$  $\mathrm{pb}^{-1}$ of $e^{+}e^{-}$ collision data collected at a
center-of-mass energy of 4170 MeV. We find no evidence of a signal, and
set an upper limit on the
branching fraction of $\mathcal{B}(\dspa\to\omega e^+\nu)<$0.20\% at the 90\% confidence level. 
 
\end{abstract}

\pacs{13.20.Fc}
\maketitle
\section{Introduction}
Multiple observations of exotic charmonium states~$[1-4]$ have been widely
interpreted as four-quark states~$[5-13]$. 
In this analysis we probe the four-quark 
content of the $\dspa$ by searching for the decay $\dspa\to\omega e^+\nu$
(charge conjugate states are implied throughout the article). 
Assuming that the $\omega$ is a pure two-quark state, its valence 
quarks are distinct
from those of the $\dspa$, and the decay can proceed through the diagram
of Fig.~1, where the $(u\bar{u})$ or $(d\bar{d})$ come from
within the $\dspa$. The initial
valence quarks annihilate while
a lepton pair is produced. Neither Cabibbo-favored, nor Cabibbo-suppressed 
decays 
can contribute to this final state. The study of this specific
process was first suggested in Ref.~\cite{bonvi}, and Ref.~\cite{gabbiani}
estimates the theoretical branching fraction for the analogous
decay $B^+\to J/\psi\ \ell^+\nu$.

\begin{figure}[htb]
\begin{center}
\includegraphics[width=4.0in]{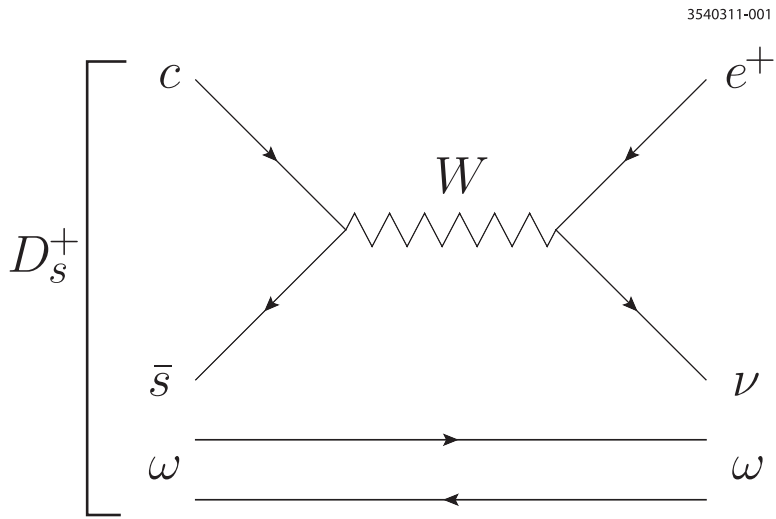}
\caption{ Feynman diagram representing the four-quark semileptonic decay $ \dspa\to \omega e^{+}\nu$.}
\label{fig:psilnu}
\end{center}
\end{figure}

Recent work by Gronau and Rosner~\cite{rosner} concludes that any value of the 
branching fraction for $\dspa \to \omega e^+ \nu$ exceeding $2 \times 
10^{-4}$ is unlikely to be explainable by $\omega$-$\phi$ mixing and would 
provide evidence for non-perturbative effects known as ``weak 
annihilation'' (see Ref.~\cite{rosner} for references).  
An estimate based on comparing 
hadronic and semileptonic processes gives a branching fraction of $(0.13 \pm 
0.05)\%$.  
This is the first search for this decay mode.

We search for a positron candidate and an 
$\omega\to \pi^+\pi^-\pi^0$ candidate, which is the
dominant decay mode with a branching fraction of 89.2\%~\cite{pdb}. 
Cabibbo-favored decays exist in the same final state, $\dspa\to\eta e^+\nu$ and
$\dspa\to\phi e^+\nu$,  
with $\mathcal{B}(\eta\to\pi^+\pi^-\pi^0)=$22.73\% and
and $\mathcal{B}(\phi\to\pi^+\pi^-\pi^0)=$15.32\%~\cite{pdb}. 
They can play the role of control samples, which are used directly
in the analysis in a variety of ways. For example, the effect of certain 
selection requirements
can be readily estimated from any change in the $\eta$ and 
$\phi$ populations. 
The two control samples are also well measured using the independent final
states $\eta\to \gamma\gamma$ and $\phi\to K^+K^-$. 
Therefore, this analysis
has good statistical sensitivity, 
and unusually strong control samples using CLEO-c data
directly. 

The remainder of this
article is organized as follows.
In Sec.~II, the detector, data, and Monte Carlo (MC) samples are described. The 
data analysis method is described in Sec.~III. The fitting procedure
is described in Sec.~IV. The determination of the branching fraction
is discussed in Sec.~V.

\section{Detector, data and MC samples}
The data used in the analysis are $e^{+}e^{-}$ collisions at a
center-of-mass energy $\sqrt{s}=$4170 MeV.
The data are collected by the CLEO-c detector and
correspond to an integrated luminosity of $586~\mathrm{pb}^{-1}$, or
$0.6\times 10^6$ $\dspa\dsma$ inclusive pairs.

The CLEO-c detector is optimized for physics in the charmonium region
and is described in detail in Ref.~\cite{cleonim}. The tracking system
consists of a central, low mass drift chamber, which
wraps directly around the beam pipe, and a main drift chamber
inside a solenoidal
magnetic field. The
particle identification 
system combines the track information from the gas chambers ($dE/dx$) and
the associated ring-imaging Cherenkov detector (RICH) data. 
The electromagnetic calorimeter consists
of CsI crystals, arranged in cylindrical fashion around the
drift chamber, to make a barrel at angles given by 
$|\cos{\theta}|<0.7$, with $\theta$ being the angle measured from the
interaction point (IP) with  
respect to the beam axis. The ends of the cylinder are
also instrumented with CsI crystals and are referred to as the end-cap regions.

MC simulations of the known physics processes~\cite{EVTGEN} 
and of the CLEO-c detector~\cite{GEANT4} are used to estimate
backgrounds and calculate signal resolution and  efficiencies.
The known charm physics processes are included in the $c\bar{c}$ MC
simulation.
All types of charm backgrounds, dominant in
this analysis, are simulated to 20 times the statistics in the data,
while the continuum $(u,d,s)$ backgrounds are simulated to 6.6 times 
the statistics in the data. 
In the following, where MC results are presented,
we multiply the continuum MC sample by 3 to obtain a consistent
$\times 20$ normalization. By convention, $c\bar{c}$ 
MC refers to the charm part
of the MC, continuum MC is the non-charm part, and MC is the weighted sum
of the two. 

The signal MC sample consists of 
8$\times 10^5$ $\dspa\to\omega e^+ \nu$ events
generated according to phase space distribution. The same number of events 
was 
generated for the $\dspa\to\eta e^+\nu$ and $\dspa\to\phi e^+\nu$ control samples.

Further samples for dominant sources of backgrounds were generated.
A sample of 2$\times 10^5$ $\dspa\to\omega\pi^+\pi^0$ events 
was separately generated,
corresponding to
about 23.5 times the number of such expected decays in the data. 
In addition, 5000 events (96 times the
data) were generated for the decay chain $\dspa\to\eta^{\prime} e^+\nu$ 
followed by
$\eta^{\prime}\to \omega\gamma$. 

\section{Data selection}

About 95\% of the $e^+e^- \to \dspa X$ events are formed through the 
following exclusive reactions~\cite{cross}
\begin{equation}
e^+e^- \to \dspa D_{\mathrm s}^{\star -}\;{\rm and}\;e^+e^- \to\dsma  D_{\mathrm s}^{\star +}.
\label{eq:reactions}
\end{equation} 
Equal amounts of positive and negative $D_{\mathrm s}^{\star}$ 
states are produced, and
about 95\% of them decay to $D_{\mathrm s}\gamma$. The analysis
described here selects exclusively both final states 
in Eq.~(\ref{eq:reactions}). In the following,
we retain both events where the $\gamma$ is associated with the positive
side, the signal side due to the convention established in Sec.~I, 
and where the $\gamma$ is associated with the negative
side, or tag side. Several kinematic constraints are available, but only
those that select both reactions in Eq.~(\ref{eq:reactions}) are used.

First we search
for an exclusively reconstructed hadronic
$\dsma$ candidate, the tag, 
and a photon candidate. 
Requiring both of these objects, with three kinematic
selections applied, strongly suppresses the backgrounds.
The photon candidate, and all tracks and showers which are 
daughters of the tag candidate,
are not used
when searching the rest of the event, which, 
due to the exclusive nature of the analysis, 
must be the signal candidate. 
The following are required to form a
signal candidate: a positron of opposite charge to the tag, precisely
three charged tracks, a $\pi^{0}$ candidate, and net event charge
equal to zero. Furthermore, the missing energy and momentum are required
to be in a relation consistent with the presence of a nearly massless
neutrino. Extra showers in the event
are ignored. 

The selection is described in detail in the remainder of this section.

\subsection{Charged and neutral particles selection}
The criteria for a good track include the requirements 
that its minimum distance to the 
IP does not exceed 5~mm in the plane perpendicular to the beam axis and 5~cm
along the beam axis. Phase space is limited to $|\cos{\theta}|<0.93$,
with $\theta$ being the angle with the beam axis,
and momentum $0.05\gev\ <p<2$~GeV.
Good tracks are then
selected as pion or kaon using dE/dx and RICH data according to the
algorithms described in detail in Ref.~\cite{briere}.

Photon candidates are contiguous groups of crystals recording significant
energy deposition. They are required to be unmatched to tracks or 
noisy crystals, 
and to have a transverse profile consistent with
expectations from an electromagnetic shower. The minimum cluster energy 
is 30 MeV. 

Positrons are selected by requiring a good track, 
and the combined positron
probability of the particle ID system and $E/p$ (the ratio 
between the shower energy associated with the candidate electron
and its track momentum) 
to be greater than 0.8.
Positron phase space requirements are  $|\cos{\theta}|<0.9$ and
$p>0.2$ GeV.

$\pi^0$ and $\eta$ 
candidates are selected by requiring a photon candidate in the 
barrel or end-cap, with $|\cos{\theta}|<0.93$. This photon is combined with a second photon, which must not be
associated with a noisy crystal, and the invariant mass of the two
photons must be within 3 standard deviations of the nominal $\pi^0$
and $\eta$ masses~\cite{pdb}.

$K^{0}_{\mathrm S}$ candidates are selected by requiring two oppositely 
charged 
tracks. If they are assigned the nominal $\pi^+$ mass~\cite{pdb}, their
invariant mass must be
within 12 MeV of the nominal $K^{0}_{\mathrm S}$ mass~\cite{pdb}.
A common vertex is calculated, and it is required to 
be radially displaced from the IP by at 
least 3 standard deviations.
\begin{table}[htb]
\caption{Definitions of $M_{\mathrm{tag}}$ signal and
sideband definitions for the tag modes.}
\begin{center}
\label{tab:modes}
\begin{tabular}{l c c c}
\hline
\hline
Mode & Signal region (GeV) &  Low sideband (GeV) & High sideband (GeV) \\
\hline

$K^0_{\mathrm S} K^- $ &[1.954, 1.983]&[1.910, 1.939] &[1.998, 2.026] \\
$K^+ K^- \pi^-$ &[1.954, 1.982] & [1.911, 1.940] & [1.996, 2.025] \\
$K^{\star -}\overline{K}^{\star 0} $&[1.953, 1.983]&[1.909,  1.938] &[1.997, 2.027] \\
$\pi^+\pi^-\pi^-$ &[1.955, 1.982] & [1.913, 1.941] & [1.996, 2.024] \\
$\eta \pi^-$ &[1.940, 2.001] & [1.892, 1.922] & [2.019, 2.050] \\
$\eta \rho^-$ &[1.940, 1.998] & [1.885, 1.914] & [2.021, 2.050] \\
$\pi^-\eta^{\prime} (\eta\pi^+\pi^-) $ &[1.944, 1.992]&[1.885, 1.933] &[2.004, 2.052] \\
$\pi^-\eta^{\prime}(\rho\gamma) $ &[1.944, 1.992]&[1.886, 1.930] &[2.002, 2.047] \\
\hline
\hline
\end{tabular}
\end{center}
\end{table}

\subsection{Tag-candidate selection}

Eight tag decay modes are used and listed in Table~\ref{tab:modes}, 
using the particle candidates selected according to
Sec.~III.A.
In addition, there are several mode-specific criteria.
For $\dsma \to K^- K^+ \pi^-$ and $\dsma \to \pi^+ \pi^- \pi^-$, the pion 
momenta are required to be greater 
than 0.1 GeV.
For $\dsma \to K^{\star -} \overline{K}^{\star 0}$, only the $(K^{0}_{\mathrm S}\pi^-)(K^+\pi^-)$
channel is considered. The $K^{\star -}$ and $\overline{K}^{\star 0}$ candidate masses are required to be
within 100 MeV of the nominal value~\cite{pdb}.
For the $\dsma \to \eta  \rho^{-} (\rho^-\to \pi^-\pi^0)$, 
the $\rho^-$ mass must be within $150$~MeV of the nominal value~\cite{pdb}.
For 
$\dsma \to \pi^- \eta^{\prime} (\eta^{\prime} \to \rho \gamma), \rho^0\to\pi^+\pi^-$, 
the $\eta^{\prime}$ mass must be within $20$~MeV of the nominal value~\cite{pdb}.
Furthermore,
the $(\pi$-$\eta^{\prime})$ helicity angle (defined as the angle $\theta_H$, in the
rest frame of the $\rho$, between the momentum of the $\pi^-$
and the momentum of the $\dsma$) is required to 
satisfy $|\cos{\theta_H}|<0.8$.

The four-momentum of a tag candidate is defined by $(E_{\mathrm tag},\mathbf{p}_{\mathrm tag})$, with the tag mass defined by $M_{\mathrm{tag}}$.
The selection further makes use of the 
recoil mass $\mreca$,
defined as
\begin{equation}
\mreca=\sqrt{(E_{\mathrm b}-E_{\mathrm{tag}})^2-(\mathbf{p}_{\mathrm b}-\mathbf{p}_{\mathrm{tag}})^2}.
\label{eq:mrec}
\end{equation} 
Here, $(E_{\mathrm b},\mathbf{p}_{\mathrm b})$ is the four-momentum 
of the colliding beams. The $\mreca$ distribution 
will peak only for those
events where the photon is associated with the signal side, but even
when the photon is associated with the tag side, 
$\mreca$ is kinematically constrained so that 
$|\mreca -M^{\star}|<$~55~MeV, where 
$M^{\star}$ is the nominal $D^{\star}_{\mathrm{s}}$ mass~\cite{pdb}.
Only candidates passing this selection are retained.  

The main tag selection is obtained from a 2-D fit described below. Because
of the complexity of the fit, the $M_{\mathrm{tag}}$ projection is fitted
first, and the fit results used to constrain some of
the final 2-D fit nuisance parameters. The
$M_{\mathrm{tag}}$ projection is also best suited for 
side-band background subtraction.

The $M_{\mathrm{tag}}$ distribution is fitted with a double Gaussian function,
$G_2$, multiplied by the fitted number of events, $N$,  and a 
first degree polynomial, $A_1$, to describe signal and background
\begin{equation}
f(M_{\mathrm{tag}})=NG_2(M_{\mathrm{tag}})+A_1(M_{\mathrm{tag}}).
\label{eq:fitg2}
\end{equation}
$G_2$ is a probability distribution composed of two Gaussian functions 
$G(x;\sigma,\mu)$,
of unit area, peaking at $\mu$ and with width equal to $\sigma$.
Here, the peak is set at $\mads$, which is the nominal
nominal $\dspa$ mass~\cite{pdb}, and the two Gaussians
have fractional probabilities $f_1$ and $(1-f_1)$
\begin{equation}
G_2(M_{\mathrm{tag}})= f_1G(M_{\mathrm{tag}};\sigma_1,\mads)+(1-f_1)G(M_{\mathrm{tag}};\sigma_2,\mads).
\label{eq:defg2}
\end{equation}

The quantities $\sigma_1$ and $\sigma_2$ are fixed to the value obtained from
the fit to the signal MC data. Having obtained $f_1$ from the fit, we
construct the variable $\sigma^2_{12}=f_1\sigma_1^2+(1-f_1)\sigma_2^2$.
The signal regions are required to be within
2.5$\sigma_{12}$ from the peak position 
for each mode except for the $(\eta \rho)$ mode where it is selected
within 2$\sigma_{12}$. 

\begin{figure}[htb]
\begin{center}
\includegraphics[width=6in]{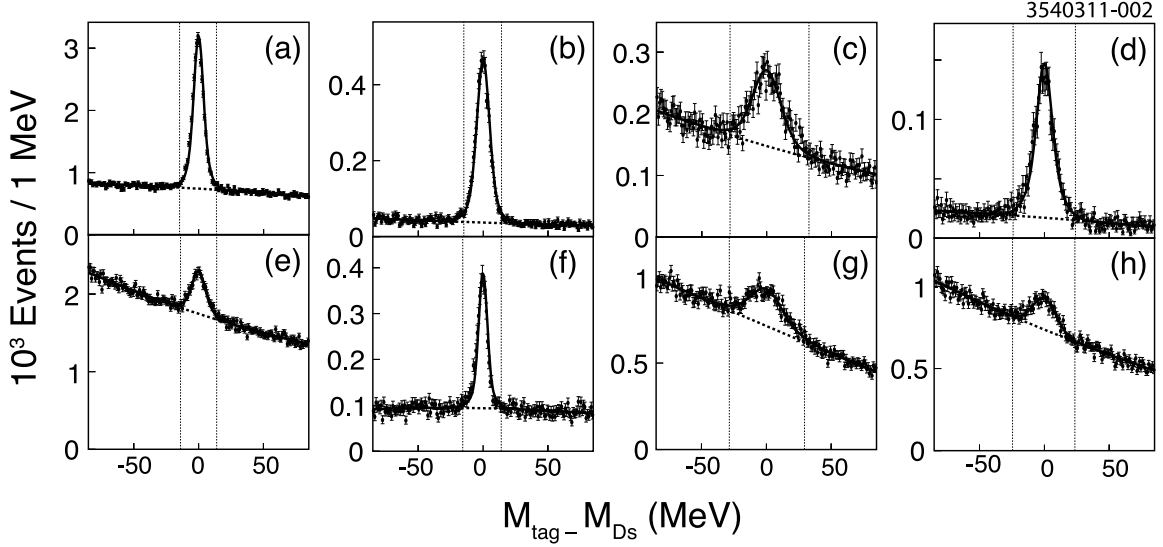}
\caption{Distribution of $M_{\mathrm{tag}}-\mads$ of
$\dsma$ candidates, for the different tags: 
(a) $K^+ K^- \pi^-$; (b) $K^0_{\mathrm S} K^- $; (c) $\eta \pi^-$; 
(d) $\pi^-\eta^{\prime} (\eta\pi^+\pi^-) $; (e) $\pi^+\pi^-\pi^-$; 
(f) $K^{\star-}\overline{K}^{\star 0} $; (g) $\eta \rho^-$, and
(h) $\pi^-\eta^{\prime}(\rho\gamma) $. The fitted background $A_1$, described
in the text, is indicated by the dashed-dotted slope. The
 signal mass region is
indicated by the vertical dotted lines.}
\label{fig:tagall}
\end{center}
\end{figure}

The $M_{\mathrm{tag}}$ data fit results are listed in 
Table~\ref{tab:tagfit}. 
The rest of the peak fit parameters are listed in Table~\ref{tab:tagfitdg}.
The $M_{\mathrm{tag}}$ distributions for the 8 modes 
are shown in Fig.~\ref{fig:tagall}.
The sidebands in the $M_{\mathrm{tag}}$ distribution 
are listed, for each mode, 
in Table~\ref{tab:modes}. 

Having determined the fit parameters for the $M_{\mathrm{tag}}$ distributions, 
a second kinematic constraint can be imposed using the $MM^{*2}$ variable
defined as

\begin{equation}
\label{eq:mmstar2}
MM^{*2}=(E_{\mathrm b}-E_{\mathrm{tag}}-E_{\gamma})^2-(\mathbf{p}_{\mathrm b}-\mathbf{p}_{\mathrm{tag}}-
\mathbf{p}_{\gamma})^2,
\end{equation} 
where  $(E_\gamma,\mathbf{p}_\gamma)$ 
is the photon four-momentum.
If the final state is given by
Eq.~(\ref{eq:reactions}), 
$MM^{*2}$ should peak at $\mads^2$.
The $MM^{*2}$ mass selection criteria are found by a two-dimensional
(2-D) binned likelihood 
fit in the ($MM^{*2},M_{\mathrm{tag}}$)
space. Each variable is also kinematically fitted, so 
that $M_{\mathrm{tag}}$ is the value obtained by constraining $MM^{*2}$ to 
its nominal value, and vice versa. This procedure improves
the signal and also minimizes any 
correlation between the two variables.

\begin{table}[htb]
\caption{Number of $D_{s}^{-}$ tag candidates $N_{\mathrm{data}}$ for each mode in
the signal region and sidebands for the one-dimensional fit to
$M_{\mathrm{tag}}$.}
\begin{center}
\label{tab:tagfit}
\begin{tabular}{l c c c}
\hline
\hline
Modes & $N_{data}$ &  Low Sideband & High Sideband \\
\hline
$K^0_{\mathrm S} K^- $ &$5828\pm 92$&1231&958 \\
$K^+ K^- \pi^-$ &$25990\pm 285$&22385&19452 \\
$K^{\star-}\overline{K}^{\star 0} $&$2891\pm 100$&2783&2647\\
$\pi^+\pi^-\pi^-$ &$8152\pm 369$&56530&43475\\
$\eta \pi^-$ &$3635\pm 160$&5727&3379\\
$\eta \rho^-$ &$6877\pm 330$&26879&14658\\
$\pi^-\eta^{\prime} (\eta\pi^+\pi^-) $&$2344\pm 70$&1040&572\\
$\pi^-\eta^{\prime}(\rho\gamma) $&$4451\pm 337$&42412&25476\\
\hline
\hline
\end{tabular}
\end{center}
\end{table}

\begin{table}[htb]
\caption{Signal peak parameters for each tag mode derived
from $c\bar{c}$ MC simulation.}
\begin{center}
\label{tab:tagfitdg}

\begin{tabular}{ l c c c }
\hline
\hline
  Mode&$f_1$&$\sigma_1$ (MeV)&$\sigma_2$ (MeV) \\
  \hline 
   $K^{0}_{\mathrm S} K^-$& 0.471&   4.05& 7.00 \\
  $K^+K^-\pi^-$& 0.725& 3.74& 8.92 \\
   $K^{\star -} \overline{K}^{\star 0}$&  0.771&  3.43& 10.65 \\
   $\pi^+ \pi^- \pi^-$&  0.899&   4.84&  9.88 \\
   $\eta \pi^-$&  0.650&   9.85& 15.56  \\
   $\eta \rho^-$&   0.574&   10.8& 18.3 \\
  $\pi^-\eta^{\prime} (\eta\pi^+\pi^-) $ &  0.590&  5.71&  13.34 \\
   $ \pi^- \eta^\prime (\rho \gamma)$&  -&   9.60&  - \\

\hline
\hline
\end{tabular}
\end{center}
\end{table}

The 2-D fit is done for each mode separately, and its purpose
is to extract the final number of tags for each mode, $N_{i}$,
while building on the information obtained in the 
one-dimensional $M_{\mathrm{tag}}$ fit. The fitting function is
\begin{equation}
\label{eq:2dfit}
f(M_{\mathrm{tag}},MM^{*2})= N_{i}G_2(M_{\mathrm{tag}})C(MM^{*2})+G_2(M_{\mathrm{tag}})A_5(MM^{*2})
+A_1(M_{\mathrm{tag}})A_5(MM^{*2}).
\end{equation}

For each mode, the signal is described by the product of a double
Gaussian in the $M_{\mathrm{tag}}$ projection, defined in Eq.~(\ref{eq:defg2}) 
and Crystal Ball function~\cite{oreglia}, defined as $C$ in the equations, 
in the $MM^{*2}$ projection, respectively. 
One of the background components is the combination of 
a real tag with a random $\gamma$. 
This type of background ($BG_1$ in Fig. 3 below,
and the second term in Eq.~(\ref{eq:2dfit})) is 
described by the same double Gaussian $G_2(M_{\mathrm{tag}})$ and a 5th 
degree polynomial $A_5(MM^{*2})$.
The other background ($BG_2$ in Fig. 3 below,
and the third term in Eq.~(\ref{eq:2dfit})) is due to 
fake tags. The PDF here is 
the product of a 1st order polynomial ($M_{\mathrm{tag}}$) and a 5th 
order polynomial ($MM^{*2}$). 
To simplify the fit, the $M_{\mathrm{tag}}$ projections are fitted 
using the signal function obtained in 
the 1-D $M_{\mathrm{tag}}$ fit, but the background parameters are varied.
\begin{figure}[htb]
\begin{center}
\includegraphics[width=6in]{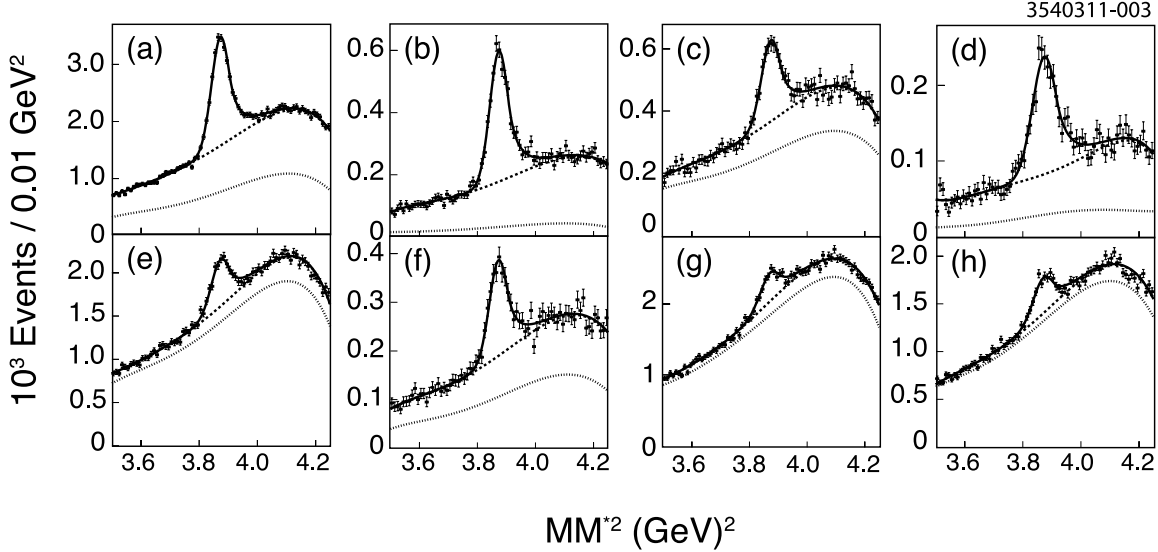}
\caption{ $MM^{*2}$ distributions for the 8 tag modes. Dash-dotted 
lines are the $BG_2$ background described in the text. Dashed lines are 
total background $BG_1+BG_2$. (a) $K^+ K^- \pi^-$; (b) $K^0_{\mathrm S} K^- $; (c) $\eta \pi^-$; 
(d) $\pi^-\eta^{\prime} (\eta\pi^+\pi^-) $; (e) $\pi^+\pi^-\pi^-$; 
(f) $K^{\star-}\overline{K}^{\star 0} $; (g) $\eta \rho^-$, and
(h) $\pi^-\eta^{\prime}(\rho\gamma) $.}
\label{fig:mmstar2fig}
\end{center}
\end{figure}

\begin{table}[htb]
\caption{$MM^{*2}$ selection range and the number of tags in each
mode obtained from the two-dimensional fit. Total number of tags,
$N_{\mathrm{tag}}$ is also given. The quoted error is statistical only.
}
\begin{center}
\label{tab:mmstar2cuts}
\begin{tabular}{l c c c }
\hline
\hline
Modes & Lower limit (\gev$^2$) & Upper limit (\gev$^2$)&$N_i$(data)\\
\hline

$K^0_{\mathrm s} K^- $ & 3.7876 & 3.9539& $3442\pm 138$\\
$K^+ K^- \pi^-$ &3.7939 & 3.9510& $15647\pm 271$ \\
$K^{\star-}\overline{K}^{\star 0} $&3.7505 & 3.9847&$1707\pm 94$ \\
$\pi^+\pi^-\pi^-$ &3.7701 & 3.9633 &$4595\pm 298$\\
$\eta \pi^-$ &3.7662 & 3.9798 &$2355\pm 187$ \\
$\eta \rho^-$ &3.7698 & 3.9632 &$3606\pm 640$\\
$\pi^-\eta^{\prime} (\eta\pi^+\pi^-) $&3.7409 & 3.9888 &$1716\pm 142$\\
$\pi^-\eta^{\prime}(\rho\gamma) $&3.7875 & 3.9601&$3373\pm 240$ \\
\hline
$N_{\mathrm{tag}}$ &-&-&$36441\pm 852$\\
\hline
\hline
\end{tabular}
\end{center}
\end{table}
The $MM^{*2}$ distributions are shown 
in Fig.~\ref{fig:mmstar2fig}. 
The $MM^{*2}$ signal regions for each mode 
are chosen so as to have 95\% signal efficiency.  
The $MM^{*2}$ selection is 
summarised in Table~\ref{tab:mmstar2cuts}. 
Table~\ref{tab:mmstar2cuts} also lists the final number of tags 
obtained in each tag mode,
$N_i$, as well as the total number of tags, $N_{\mathrm{tag}}$, used to extract
the final result. 

\subsection{Signal selection}

The signal is selected by requiring one positron candidate, of charge 
opposite to the tag charge, 
two charged pion candidates, of opposite charge,
no extra good tracks, and a good $\pi^0$, all selected
exclusively of the objects used in the tag. The selection requires 
a specific number of tracks, and multiple 
candidates can arise only due to multiple $\pi^0$ candidates. 
In case of multiple
candidates, the $\pi^0$ is selected as follows. Given the photon-photon mass
$M_{\gamma\gamma}$, and the calculated mass error $\sigma_{\gamma\gamma}$,
the one with the lowest 
$\chi^2=[(M_{\gamma\gamma}-M_{\pi^0})/\sigma_{\gamma\gamma}]^2$ is chosen. 
Additional
candidate photons are ignored.

\begin{figure}[htb]
\begin{center}
\includegraphics[width=6in]{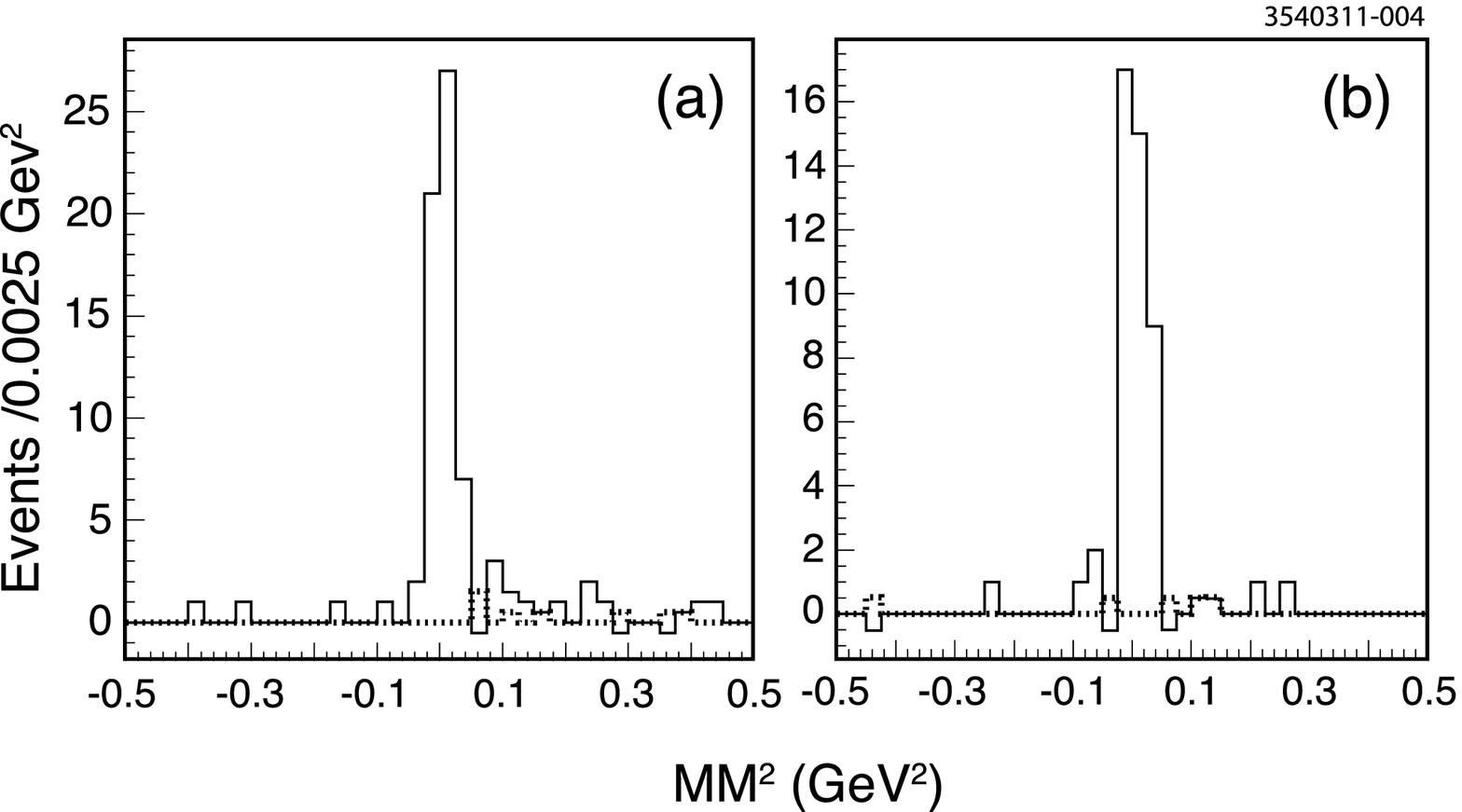}
\caption{Control sample $MM^2$ distributions. 
(a) solid, $\dspa\to\eta e^+\nu$ distribution after $M_{\mathrm{tag}}$
sideband subtraction, and  $M_{\mathrm{tag}}$
sideband distribution (dotted). (b)
solid, $\dspa\to\eta e^+\nu$ distribution after $M_{\mathrm{tag}}$
sideband subtraction, and  $M_{\mathrm{tag}}$
sideband distribution (dotted).
}
\label{fig:mm2eta}
\end{center}
\end{figure}

The positron, charged pions, and $\pi^0$ are added together to form the
four-vector ($E_{\mathrm s},\mathbf{p}_{\mathrm s}$). The measured neutrino candidate mass 
squared, $MM^2$, is defined as
\[ MM^{2}=(E_{\mathrm b}-E_{\mathrm{tag}}-E_{\gamma}-E_{\mathrm s})^2-(\mathbf{p}_{\mathrm b}-\mathbf{p}_{\mathrm{tag}}-
\mathbf{p}_{\gamma}-\mathbf{p}_{\mathrm s})^2.\] 
The $MM^2$ distributions, with  $M_{\mathrm{tag}}$ sideband subtraction,
of the two control samples $\eta e^+\nu$ and $\phi e^+\nu$
are shown in Fig.~\ref{fig:mm2eta}. 
Based on the shape of $MM^2$, events with $-0.05\gev^2<MM^2<0.05$
~GeV$^2$ are selected for the final analysis.

\begin{figure}[htb]
\begin{center}
\includegraphics[width=6in]{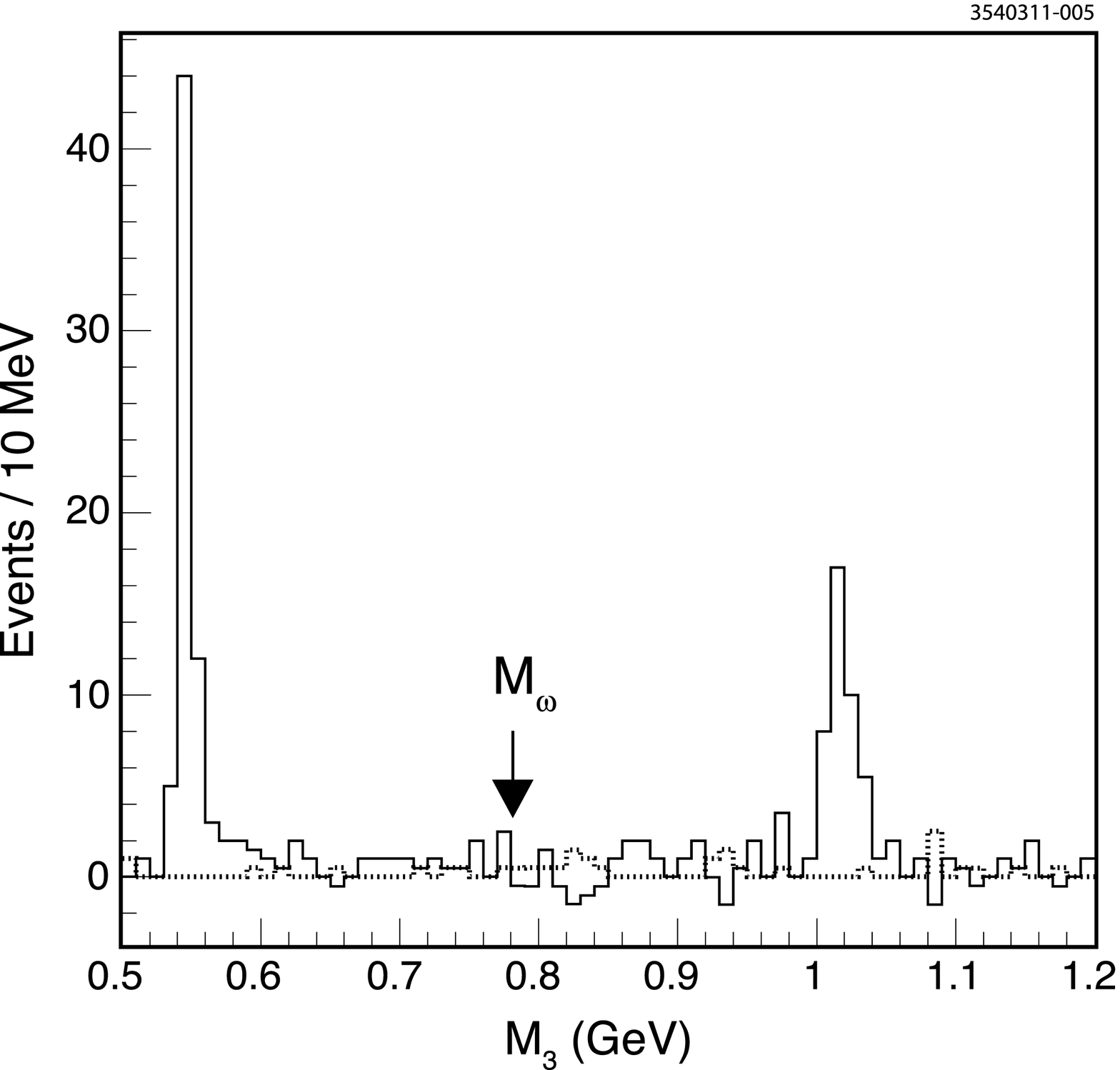}
\caption{$M_3$ distribution. Solid: 
signal selection, after $M_{\mathrm{tag}}$ sideband subtraction. 
Dotted: $M_{\mathrm{tag}}$ sideband contribution. The arrow shows the
location of the $\omega$ nominal mass, Ref.~$[17]$.}
\label{fig:pipipi0}
\end{center}
\end{figure}
The mass of the $\pi^+\pi^-\pi^0$ combination $M_3$ 
was not used in the candidate selection, and provides
the spectrum that is fitted to extract the 
final result. In Fig.~\ref{fig:pipipi0},
the $M_3$ spectrum is presented, including
$M_{\mathrm{tag}}$ sideband contributions. 
Two peaks are clearly present, at the $\eta$ and $\phi$ masses, with no sign
of a signal in the $\omega$ mass region.

The $M_3$ peaks in the signal MC samples are fitted to a Breit-Wigner shape (indicated as $BW$ in the equations),
convoluted with a double Gaussian, 
\begin{equation}
\label{eq:defsig}
s(x)=K \int BW(x_1)G_2(x-x_1)dx_1,
\end{equation}
where $K$ is a normalization constant to give $s(x)$ a unit area. 
Table~\ref{tab:dg} lists the
fit results for each of the signal MC samples generated for this analysis. 
\begin{table}[htb]
\caption{$M_{3}$ signal peak parameters evaluated from signal
MC sample. 
All quantities are defined in the text.}\label{tab:dg}
\begin{center}
\begin{tabular}{l c c c c}
\hline
\hline
Decay & $f_1$ & $\sigma_1$ (MeV) & $\sigma_2$ (MeV) & R.M.S. (MeV) \\
\hline
 $\eta e^+\nu $ & 0.8844 & 3.165 & 19.85 & 7.37 \\ 
 $\omega e^+\nu $ & 0.8783 & 5.500 & 22.52 & 9.40 \\
 $\phi e^+\nu $ & 0.8361 & 5.940 & 19.83 & 9.73 \\
\hline\hline
\end{tabular}
\end{center}
\end{table}

The reconstruction
efficiency $\epsilon$ for the
$\omega e^+\nu$ final state is computed by applying the same requirements
to the signal 
MC events, but correcting for the number of tags found in the data, 
\begin{equation}
\label{eq:epsil}
\epsilon={1\over N_{\mathrm{tag}}}{\Sigma N_i\epsilon_i},
\end{equation}
$\epsilon_i$ being the signal MC efficiency for mode $i$.
The result is $\epsilon=(5.11\pm 0.15)\%$, with the error due to MC statistics.

\section{ Final fit}

\begin{figure}[htb]
\begin{center}
\includegraphics[width=4.0in]{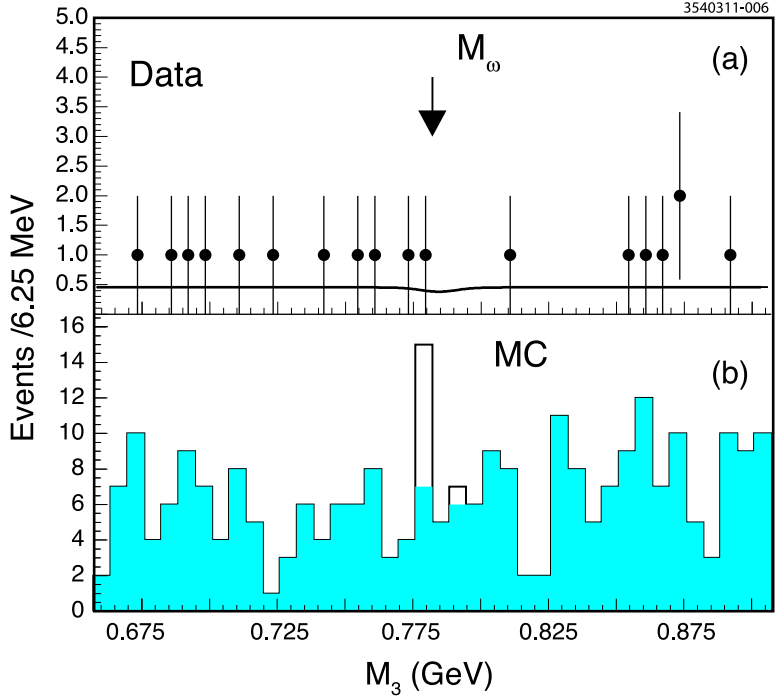}
\caption{$M_3$ distribution in the 
250 MeV wide 
region centered at the nominal $\omega$ mass. No $M_{\mathrm{tag}}$ 
sideband subtraction was used.
(a) comparison of
data (points) and best fit according to Eq.~(9) (line).
(b) The MC sample, normalized to 20 times the data statistics. 
Solid: non-resonant backgrounds. 
Empty: resonant backgrounds. The arrow shows the
location of the $\omega$ nominal mass.}
\label{fig:omegamc}
\end{center}
\end{figure}

Figure~\ref{fig:omegamc} shows only the $M_3$ region used in the fit, which
contains $\noa=18$ events. The $\Delta_{M3}=250$ MeV mass window
is centered at the nominal $\omega$ mass~\cite{pdb}.
In the Fig.~6(a), the data distribution
and the fit to the data (described below) are shown. 
In Fig.~6(b), the MC distribution is shown.

Three potential sources of background
are considered: non-$D_{\mathrm s}$ backgrounds, 
$D_{\mathrm s}$ backgrounds where there are
non-resonant final states (which have not yet been observed, and are 
not present in the MC simulation), 
and backgrounds where there is a true
$\omega$. $M_{\mathrm{tag}}$ sideband subtraction only subtracts the 
first source.
A direct fit of a signal and a background component subtracts
the first two. The third source of background 
is subtracted via MC simulation,
and is discussed below.
\begin{figure}[htb]
\begin{center}
\includegraphics[width=4.0in]{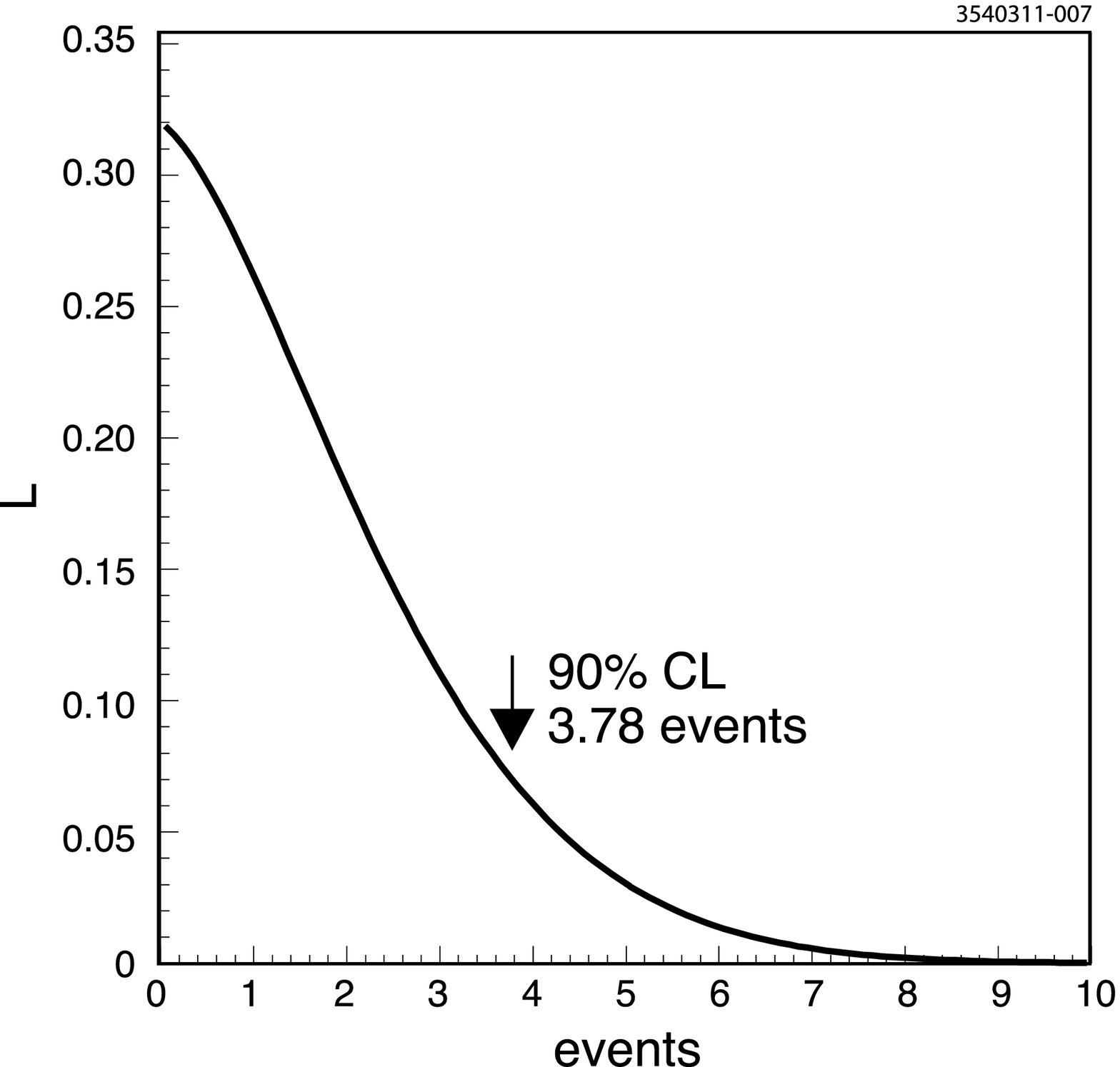}
\caption{Statistical only
likelihood, normalized to unit area over the positive signal region, 
for the observed number of signal
events in data. The 90\% C.L. was computed using only the 
positive signal region.  
}
\label{fig:likeplots}
\end{center}
\end{figure}
The signal yield is determined by a one 
parameter unbinned likelihood fit~\cite{pdb}.
The free parameter is the total number of signal events $S$. The background
level is constrained by the normalization of the probability.
$S$ is multiplied by a function of unit area $s(x)$, Eq.~(\ref{eq:defsig}).
The final 
expression of the unbinned likelihood, $L_u=\Pi_{i}P_{i}$, is obtained from
the probabilities
\begin{equation}
\label{eq:likeli}
P(M_{3i}|S)=P_i= (S/\noa)s(M_{3i})+(1-S/\noa)/\Delta_{M3},
\end{equation}
which correspond to a signal $S$, distributed according to $s(M_3)$, 
plus a flat
background.

Figure~\ref{fig:likeplots} 
shows the likelihoods obtained for data, 
without any peaking background subtraction,
in the $S>0$ region. The 90\% confidence level (C.L.) is calculated using only
the $S>0$ portion of the likelihood. The statistical only upper limit on $S$
at the 90\%  C.L. is $S_{90}=3.78$ events.

\begin{figure}[htb]
\begin{center}
\includegraphics[width=4.0in]{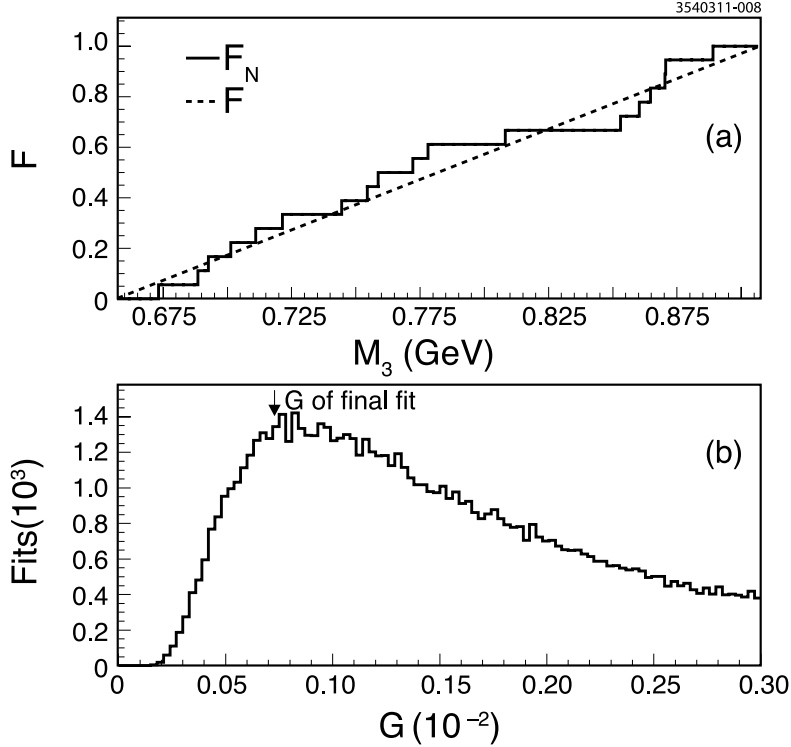}
\caption{Cramer-Von Mises test of goodness of fit, for the final fit of this
analysis. (a) Comparison of the  integrated probability 
distribution $F$ (dashed)
and the step function $F_N$ described in the text (solid). (b) Comparison of
the $G$ obtained in this analysis, with a distribution obtained from $10^5$
unbiased toy MC fits.}
\label{fig:cramer}
\end{center}
\end{figure}

Unbinned likelihood fits, in one dimension, can be tested for goodness of fit
using the Cramer-Von Mises test~\cite{cramer}, 
where the goodness of fit parameter is

\begin{equation}
\label{eq:g}
G=\int^{M_{3}^{\mathrm{max}}}_{M_{3}^{\mathrm{min}}}[F(M_3)-F_N(M_3)]^2 dF(M_3).
\end{equation}
The integral limits are the limits of the fit interval.
$F(M_3)$ is the integrated probability function for best-fit parameters, 
\begin{equation}
\label{eq:f}
F(M_3)=\int_{M_{3}^{\mathrm{min}}}^{M_3} P(M'_3,S_{\mathrm{max}}) dM'_3.
\end{equation} 
Here $S_{\mathrm{max}}=0$, so that $F$
is in fact a straight line. One has 
$F=(M_{3}-M_{3}^{\min})/\Delta_{M_{3}}$ and
$dF(M_3)=dM_{3}/\Delta_{M_{3}}$.
$F_N$ is a step function such 
that $F_N(M_3)=N/\noa$,
where $N$ is the rank of the largest event mass which is less than $M_3$.
The two functions are shown in Fig.~\ref{fig:cramer}.

A toy MC program was run to generate an
ensemble of $10^5$ unbiased experiments. Figure~\ref{fig:cramer} shows the distribution
of $G$ for the ensemble, also shown is the value of $G$ obtained in the fit
to data. Only 13.1\% of the fits to the generated experiments are better
than that to the data.
The toy MC program also made it easy to apply
the Kolmogorov-Smirnov (KS) test, which simply
computes the maximal difference between $F$ and $F_N$. 
Only 16.0\% of the unbiased experiments
produced a better  KS test than the data. The 
fit to the data is excellent.
 
\section{Determination of branching fraction and systematic errors}

The statistical upper limit on the number of events is translated into a 
statistical only limit on the 
branching fraction $\mathcal{B}_{90}$ according to the following equation
\begin{equation}
\label{eq:b95}
\mathcal{B}_{90}={S_{90}\over \epsilon N_{\mathrm{tag}}}.
\end{equation}

There are three  quantities on the right hand side of Eq.~(\ref{eq:b95}), 
with central
values $S_{90}=3.78,\ \epsilon=0.0511$ and $N_{\mathrm{tag}}=36441$,
yielding $\mathcal{B}_{90}=0.203$\%, which is a 
purely statistical limit.
$N_{\mathrm{tag}}$ has a statistical error of its own, and
each of the three quantities in Eq.~(\ref{eq:b95})
has systematic errors which are 
discussed below.

\subsection{Systematic errors}

Table~\ref{tab:sys}, first row, contains the relevant parameters of the 
unbinned 
likelihood (Fig.~\ref{fig:likeplots}), 
in the form $\mu\pm\sigma$. $\mu$ is the
$S$ value for which $L_u$ is maximal,
if one allows also $S<0$ values. It describes the form of the likelihood, but
is not used in the determination of the final result.

Systematic errors to $S_{90}$ are also listed in Table~\ref{tab:sys}. 
The error associated with the assumed
mass and width of $\omega$ is estimated by varying the central values by
the uncertainties given in Ref.~\cite{pdb}.

The greatest source of $S_{90}$
systematic errors is related to irreducible backgrounds.
These also shift the location of the likelihood
peak to lower values.
Fig.~\ref{fig:omegamc} shows the background distribution by physical source.
All but one of the true $\omega$  
are due to the decay chain $\dspa\to\eta^{\prime} e^+\nu$ with
$[\mathcal{B}=(1.12\pm 0.35)\%]$, followed by
$\eta^{\prime}\to \omega\gamma$, with 
$[\mathcal{B}=(3.02\pm 0.33)\%]$~\cite{pdb}. 
A dedicated MC simulation for this channel generated 5000 events of which 
51 passed all selections. This 
corresponds to an irreducible background of $(0.53\pm 0.19)$ events. 
Note that
the largest source of error is the semileptonic $\mathcal{B}$ error 
from Ref.~\cite{pdb}. Therefore,
this source of systematics can not be significantly improved with increased
simulation statistics.

A second irreducible
background comes from $D_{\mathrm s}\to\omega X$ events. 
Zero events are found in the MC events from the direct decay $\dspa\to\omega \pi^+$. 
The $c\bar{c}$ MC significantly underestimates the $(D_{\mathrm s}\to\omega X)$ yield, which
is 0.6\% in the $c\bar{c}$ MC but 6.1\% in data~\cite{roches}. The decay
$\dspa\to\omega\pi^+$ is in the $c\bar{c}$ MC, but no other $\omega n(\pi)$ decays.
A dedicated MC for $\dspa\to\omega\pi^+\pi^0$, which was assumed to
saturate the 5.5\% difference, was run, and zero events were found. The 
probability for $n$ background events, given zero MC candidates, is 
exponential in shape. The systematic error from this source 
can be represented as $(0.02\pm 0.02)$ in 
Table~\ref{tab:sys}. 

There was one more true $\omega$ event
which is in the continuum MC sample,
corresponding to one more irreducible background of $0.15\pm 0.15$ events.

\begin{table}[htb]
\caption{Summary of statistical and systematic errors of $S_{90}$. The first block is the statistical error from the experiment. The second block are errors 
associated with the quantity $S_{90}$ of
Eq. (12) and consists of uncertainties due to the mass and width of the
omega, and errors in estimating three irreducible backgrounds as
described in the text. The last block are percentage
systematic errors
associated with $N_{\mathrm{tag}}$ or $\epsilon$.}
\begin{center}
\label{tab:sys}
\begin{tabular}{l c c }
\hline
\hline
Type & Cent. val. (evts.) & $\sigma$ (evts.)  \\
\hline
Data fit & $-0.25$ & 2.21 \\
\hline
$\omega$ mass & $-$ & 0.04 \\
$\omega$ width & $-$ & 0.006  \\
$\dspa\to \eta^{\prime} e^+\nu$ & $-$0.53 & 0.19    \\
$\dspa\to \omega X$ & $-$0.02 & 0.02  \\
Continuum & $-$0.15 & 0.15  \\
\hline
Type & Cent. val.(\%) & $\sigma$ (\%)  \\
$N_{\mathrm{tag}}$ stat.& $-$ & 2.3 \\
$N_{\mathrm{tag}}$ syst.& $-$ & 2.0 \\
MC statistics & $-$ & 2.7  \\
 $\mathcal{B}(\omega\to 3\pi)$ & $-$ & 0.8   \\ 
Tracking & $-$ & 0.9  \\
$\pi^0$ eff. & $-$ & 1.0  \\
$\pi^0$ selection variation & $-$ & 0.5  \\
Positron eff. & $-$ & 0.6 \\
MC form factor &$-$ & 0.5 \\
Extra track selection &$-$ & 0.04  \\
Particle ID &$-$ & 0.1  \\
\hline\hline
\end{tabular}
\end{center}
\end{table}

$N_{\mathrm{tag}}$ was obtained through a fit, with a statistical error of 2.3\%.
Systematic errors can enter 
the analysis only through the bias in the choice of fitting function. This
can be quantified by varying the fitting function.
For each tag mode, the fitting function for the signal was changed,
first term of Eq.~(\ref{eq:2dfit}). 
The Crystal Ball function was varied in two
ways, by keeping the $n$ parameter fixed to its MC fitted values
and by changing the $(n,\alpha)$ parameters by one $\sigma$
in a mode specific way.
The background was also varied. Instead of a fifth degree polynomial,
the data were fitted with a fourth and a sixth degree polynomial.
The background was also changed by fixing the amount of $BG_1$ background
(described in Sec.~III.B) 
to one standard deviation above or below its central value.
Variations of $N_{\mathrm{tag}}$ due to changes in the fitting function were as
low as $-1.8\%$ and as high as $+1.5\%$.
The assigned systematic $N_{\mathrm{tag}}$ error is 2.0\%. 

Correlations may affect the fit of the 
$(M_{\mathrm{tag}},MM^{*2})$ peak, because the fit assumes the two variables are not 
correlated. To study this, we have computed the $(M_{\mathrm{tag}},MM^{*2})$
correlation coefficient in the signal MC sample, by calculating
the correlation coefficient in each tag mode, and then reweighting for the
observed number of events in each mode. The result is $\rho=(-1.6\pm 0.9)\%$.
The fit error due to remnant correlations
is of order $\rho^2$ and is neglected.

Finally, there are the systematic
uncertainties on the efficiency to be considered.
The $\omega$ branching fraction uncertainty is 0.8\% ~\cite{pdb}.
The tracking efficiency error is a 0.3\% Gaussian systematic error per signal
track,
to be added linearly, totaling 0.9\% per event~\cite{naik}. 

The
$\pi^0$ reconstruction efficiency error is
1\%~\cite{dobbs}, but depends on the exact selection criteria. 
To estimate the size of the systematics induced by changing selection
criteria, the signal MC sample with and without the energy and
angular criteria which were used to select photon candidates. There were 41269
reconstructed events with the criteria, and 41868 without
the criteria, a difference of 1.5\%. There were 101 events instead of 99 in the
combined $\eta$ and $\phi$ peaks, a difference of 2\%. 
We assumed a further 0.5\% systematic error, listed
in Table~\ref{tab:sys}.  
 
The positron reconstruction efficiency is evaluated in a manner
similar to Ref.~\cite{dobbs}. Positron efficiencies
have been investigated by the Collaboration using a variety
of well-known kinematically constrained QED processes. 
The experimentally measured corrections are convoluted
with the positron momentum distribution to obtain
the efficiency uncertainty, which is 0.6\%. 
The effects of the extra track cuts and of particle ID
cuts can be estimated from the MC sample, by
varying or eliminating the cuts. 
We find errors of 0.04\% and 0.1\% for the extra track cut 
and particle ID cuts respectively.

Finally,  a different form factor will change the
efficiency, mostly 
because events with low $q^2$ (the positron-neutrino mass) 
produce lower energy positrons. The signal MC produce phase-space distributed
events, and therefore a constant form factor.
To evaluate this source of systematics, the form factor was varied by 
$\pm 20\%$, by reweighting the signal MC events according to the weights

\[w_{\pm i}(q^2_i)=1\pm \frac{0.2(q^2_i-<q^2>)}{<q^2>},\]
with  $<q^2>$ being the mean $q^2$ in the signal MC sample.
The new efficiencies are 5.08\% and 5.14\% respectively, to be compared
to the given value of 5.11\%. A systematic error of 0.5\% is assigned to 
this systematics.

\begin{figure}[htb]
\begin{center}
\includegraphics[width=4.0in]{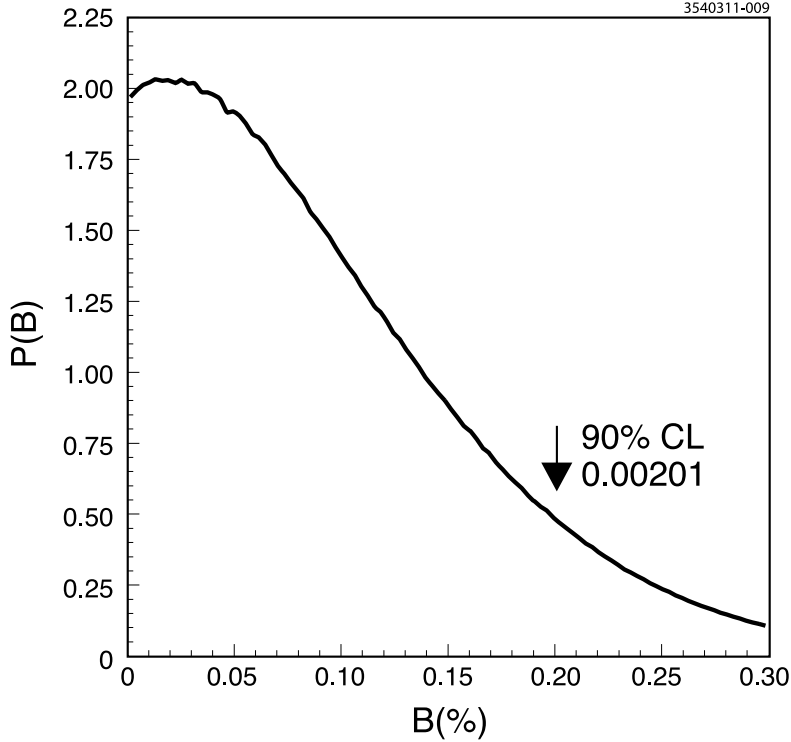}
\caption{Final probability distribution for the branching fraction, after
convolution of all statistical and systematic errors, as described in the text.
}
\label{fig:finaleff}
\end{center}
\end{figure}

\subsection{Determination of the branching fraction}
To obtain the final result, Gaussian and non-Gaussian errors 
are convoluted with the non-Gaussian 
signal $S$ distribution given by the likelihood (Fig.~\ref{fig:likeplots}), 
by means of a toy MC program. 
The procedure is used for several reasons. First, 
the main source of error, the unbinned likelihood, is non-Gaussian, whereas
the smaller sources of error are mostly Gaussian. Second,
some sources of error shift the central value of the likelihood, an effect
which can be treated exactly by shifting the likelihood on an event-by-event 
basis. Finally, the exponential and correlated nature of some of the 
error sources can be reproduced exactly by MC simulation.

A total of $25\times 10^6$ toy experiments are generated, to obtain 
the probability distribution in
Fig.~\ref{fig:finaleff}. The
cumulative effect of the systematic errors is to increase the 
limit, and the cumulative effect of the irreducible backgrounds is to
decrease the limit. Taking into account the systematic uncertainties and 
the irreducible backgrounds,
the upper limit on the branching fraction changes 
from 0.203\% (statistical only) to
0.201\%.

\section {Conclusion}

We report the first measurement of an upper limit for the branching
fraction ${\mathcal B}(\dspa\to\omega e^+\nu)$. We find 
${\mathcal B}(\dspa\to\omega e^+\nu)<$0.20\% at the 90\% C.L.,
which does not exclude that expected from
the model of Ref.~\cite{rosner}.

\begin{acknowledgments}
We gratefully acknowledge the effort of the CESR staff
in providing us with excellent luminosity and running conditions.
D.~Cronin-Hennessy thanks the A.P.~Sloan Foundation.
This work was supported by
the National Science Foundation,
the U.S. Department of Energy,
the Natural Sciences and Engineering Research Council of Canada, and
the U.K. Science and Technology Facilities Council.
\end{acknowledgments}

\nopagebreak

\end{document}